\documentclass[a4paper,12pt]{amsart}
\usepackage{natbib}
\usepackage[T1]{fontenc}
\usepackage{amsmath}
\usepackage{amssymb}
\usepackage[obeyspaces]{url}
\usepackage{soul}

\newcommand{\fF}{\ensuremath{\mathfrak{f}}}

\newcommand{\FF}{\ensuremath{\mathfrak{F}}}
\newcommand{\sF}{\ensuremath{\mathfrak{s}}}
\newcommand{\nF}{\ensuremath{\mathfrak{n}}}
\newcommand{\SSF}{\ensuremath{\mathsf{S}}}
\newcommand{\TSF}{\ensuremath{\mathsf{T}}}
\newcommand{\GSF}{\ensuremath{\mathsf{G}}}

\title{On a possible foundation of a theory of matter}

\thanks{Originally published in German as ''{\"U}ber eine m{\"o}gliche Grundlage einer Theorie der Materie'', \textit{\"Oversigt af Finska Vetenskaps-Societetens F\"orhandlingar} (Helsingfors), Bd. LVII. 1914-1915. Afd. A. N:o 28, p. 1--21. Translated by Frank Borg, present address: University of Jyv\"askyl\"a, Chydenius Institute, POB 567, 67101-Karleby, Finland; email: \url{borgbros@netti.fi}. A scanned pdf-version of the original paper along with the two other papers that Nordstr\"om published on the topic can be found at this url: \url{www.netti.fi/~borgbros/nordstrom}. For some brief historical sketches see Ch.\ Cronstr{\"o}m, ''Gunnar Nordstr{\"o}m (1881--1923)'' p.\ 3--9; S.\ Deser, ''The many dimensions of dimension'', p.\ 65--74; F.\ Ravndal, ''Scalar gravitation and extra dimensions'', p.\ 151--164: in Ch.\ Cronstr{\"o}m and C.\ Montonen (eds), \textit{Proceedings of the Gunnar  Nordstr{\"o}m Symposium on Theoretical Physics}, August 27--30, 2003, Helsinki. Commentationes Physico-Mathematicae 166/2004, The Finnish Society of Sciences and Letters.} 
\author{Gunnar Nordstr\"om}

\begin{document}

\begin{abstract}
This is the third paper by Nordstr\"om on his five dimensional theory. Here Nordstr\"om attempts to base a theory of matter on the energy conservation generalized to the five dimensional case. This involves a new fundamental constant related to the elementary electrical charge. --F.B.
\end{abstract}

\maketitle

In the two previous communications\footnote{G. \so{Nordstr{\"o}m}, Phys.\ Zeitschr.\ \textbf{15}, p.\ 504, 1914; this journal LVII.\ 1914-1915.\ A.\ N:o 4. The abbreviation l.\ c.\ will be used here to refer to the latter contribution.} have I shown how one can unite the differential equations of the electromagnetic and gravitational fields in a symmetrical fashion, when one views the four dimensional spacetime-world as a surface in a five dimensional world-extension. In the second of the cited works I indicated that the suggested point of view could be useful for setting up a theory of matter; the elaboration of this theme is the aim of the present communication. I must however from the outset emphasize that I have not succeeded in bringing these investigations to a satisfactory conclusion.

In the general case the total field encompasses two ten-vectors \fF~ and \FF~ (the actual field vectors) and two five-vectors \sF~ and $\varPhi$ (five-current and five-potential). The components of these vectors are related by the differential equations (I), (II), (1) l.\ c. Furthermore, \fF~ and \FF~ are related by some supplementary conditions about which one may make various assumptions. 

Among the state variables we have still the five dimensional tensor \SSF, which is related to the the matter-elastic tensor \TSF~ as indicated by the equations (2) l.\ c. As we will adhere to the theory of gravitation developed by me in Ann.\ d.\ Phys. 42\footnote{G. \so{Nordstr{\"o}m}, Ann.\ d.\ Phys. 42, p.\ 533, 1913.}, we must rely on the equations (27) l.\ c. \nF~ is a five-vector normal to the world-surface. We take the normal direction to be the $w$-direction so that $\nF_x = \nF_y = \nF_z = \nF_u = 0$, $\nF_w \gtrless$ 0. We have then by (27) l.\ c.\

\begin{equation}
\label{EQ:a}
\tag{a}
\left\{
\begin{aligned}
\SSF_{xw} &= - \sF_x \varPhi_w, \; \SSF_{yw} = - \sF_y \varPhi_w,
\\
\SSF_{zw} &= - \sF_z \varPhi_w, \; \SSF_{uw} = - \sF_u \varPhi_w,
\\
\SSF_{ww} &= - \sF_w \varPhi_w, 
\end{aligned}
\right.
\end{equation}

As I have remarked on p.\ 12 l.\ c.\ it lies close at hand to assume, for a theory of matter, that the tensor \SSF~ is fully determined by the vectors \sF~ and $\varPhi$\footnote{It is to be observed that the spacetime components of $\varPhi$ do not necessarily coincide with the ordinary electrodynamical potential, since we cannot maintain that

\[
\frac{\partial \varPhi_x}{\partial x} +
\frac{\partial \varPhi_y}{\partial y} +
\frac{\partial \varPhi_z}{\partial z} +
\frac{\partial \varPhi_u}{\partial u} 
\]

is equal to zero. For a \so{static} field we have however

\[
\varPhi_x = \varPhi_y = \varPhi_z = 0, \quad
\frac{\partial \varPhi_u}{\partial u} = \frac{\partial \varPhi_w}{\partial u} = 0,
\]
and thus $\varPhi_u$ and $\varPhi_w$ are ordinary scalar potentials in this special case. A similar relation holds for the theory by \so{Mie}; see G.\ \so{Mie}, Ann.\ d.\ Phys.\ \text{37}, p.\ 511, 1912.}. The equations which express \SSF~ in terms of \sF~ and $\varPhi$ must of course include the equations (\ref{EQ:a}), and hence we make the following ansatz:

\begin{equation}
\label{EQ:1}
\left\{
\begin{aligned}
\SSF_{ab} = - \sF_a \varPhi_b,
\\
\SSF_{aa} = - \sF_a \varPhi_a.
\end{aligned}
\right.
\end{equation}

When the five dimensional tensor \SSF~ is symmetric, then the two five-vectors \sF~ and $\varPhi$ must be parallel,

\begin{equation}
\label{EQ:2a}
\tag{2 a}
\sF = \beta \, \varPhi,
\end{equation}

where $\beta$ is a five dimensional scalar with the dimension $l^{-2}$.  
We may note, however, that the spacetime components of \SSF~ may very well be symmetric even if

\[
\SSF_{xw} \gtrless \SSF_{wx} \quad \mbox{etc}. 
\]

In this case we would have

\begin{equation}
\label{EQ:2}
\left\{
\begin{aligned}
\sF_x = \beta_1 \varPhi_x, \; \sF_y = \beta_1 \varPhi_y, 
\\
\sF_z = \beta_1 \varPhi_z, \; \sF_u = \beta_1 \varPhi_u,
\\
\sF_w = \beta_2 \varPhi_w,
\end{aligned}
\right.
\end{equation}

where $\beta_1 \gtrless \beta_2$. The inequality between $\beta_1$ and $\beta_2$ is possible because the direction normal to the world-surface is a distinguished direction\footnote{Moreover, one possibility is that the tensor \TSF~ is symmetric while \SSF~ is not. Then the equations (2) l.\ c.\ would have to be changed in such a way that the off-diagonal components would satisfy

\[
\frac{1}{2}(\SSF_{xy} + \SSF_{yx}) = \TSF_{xy} = \TSF_{yx} \quad \mbox{etc}.
\]

Instead we will maintain the equality between the off-diagonal components of the two tensors.}.

We have not to investigate what sort of conditions our theory impose on the momentum-energy theorem. In the five dimensional case the theorem will be expressed through five equations, one for each coordinate axis. The equation for the $u$-direction expresses the energy theorem and has the form

\begin{equation}
\label{EQ:3}
\left\{
\begin{aligned}
-\frac{\partial}{\partial x} \left\{ \GSF_{xu} + \TSF_{xu} \right\}
-\frac{\partial}{\partial y} \left\{ \GSF_{yu} + \TSF_{yu} \right\}
-\frac{\partial}{\partial z} \left\{ \GSF_{zu} + \TSF_{zu} \right\} -
\\
-\frac{\partial}{\partial u} \left\{ \GSF_{uu} + \TSF_{uu} \right\}
-\frac{\partial}{\partial w} \left\{ \GSF_{wu} + \TSF_{wu} \right\} = 0.
\end{aligned}
\right.
\end{equation}

\GSF~ is the stress-energy tensor of the unified electromagnetic and gravitational fields, \TSF~ is the matter-elastic tensor, which is related to the tensor \SSF~ according to the equations (2) l.\ c. By adding the five equations (2) l.\ c.\ for the diagonal components we find that

\[
\sum_a \SSF_{aa} = -4 \sum_a \TSF_{aa},
\]

and we have also that

\begin{equation}
\label{EQ:4}
\left\{
\begin{aligned}
\TSF_{xx} = \SSF_{xx} - \frac{1}{4} \sum_a \SSF_{aa},
\\
\TSF_{xy} = \SSF_{xy} \quad  \mbox{etc.}   
\end{aligned}
\right.
\end{equation}

It follows then from (\ref{EQ:1})

\begin{equation}
\label{EQ:5}
\left\{
\begin{aligned}
-\TSF_{xx} &= \sF_x \varPhi_x - \frac{1}{4}
\left\{
\sF_x \varPhi_x +
\sF_y \varPhi_y +
\sF_z \varPhi_z +
\sF_u \varPhi_u +
\sF_w \varPhi_w 
\right\},
\\
-\TSF_{xy} &= \sF_x \varPhi_y \quad \mbox{etc}.
\end{aligned}
\right.
\end{equation}

In order to arrive at the energy-relation from the field equations, we have to multiply the three first equations in (II) l.\ c.\ with $\FF_{ux}$, $\FF_{uy}$, $\FF_{uz}$, the fifth equation with $\FF_{uw}$, and further the second, third, fourth, eighth, ninth and tenth equations in (II) l.\ c.\ with $\fF_{yz}$, $\fF_{zx}$, $\fF_{xy}$, $\fF_{wx}$, $\fF_{wy}$, and $\fF_{wz}$ resp. By adding the resulting ten terms we obtain after some rearranging\footnote{Cmp. G. \so{Nordstr{\"o}m}, Phys.\ Zeitschr.\ \textbf{15}, p.\ 505, 1914, where the same calculation is made in the case \fF~ = \FF.},

{\allowdisplaybreaks
\begin{align*}
\FF_{ux} 
\left(
\frac{\partial \fF_{xy}}{\partial y} +
\frac{\partial \fF_{xz}}{\partial z} +
\frac{\partial \fF_{xw}}{\partial w} 
\right) +
\FF_{uy} 
\left(
\frac{\partial \fF_{yx}}{\partial x} +
\frac{\partial \fF_{yz}}{\partial z} +
\frac{\partial \fF_{yw}}{\partial w} 
\right) +
\\
\FF_{uz} 
\left(
\frac{\partial \fF_{zx}}{\partial x} +
\frac{\partial \fF_{zy}}{\partial y} +
\frac{\partial \fF_{zw}}{\partial w} 
\right) +
\FF_{uw} 
\left(
\frac{\partial \fF_{wx}}{\partial x} +
\frac{\partial \fF_{wy}}{\partial y} +
\frac{\partial \fF_{wz}}{\partial z} 
\right) +
\\
\fF_{yz} 
\left(
\frac{\partial \FF_{zu}}{\partial y} +
\frac{\partial \FF_{uy}}{\partial z} 
\right) +
\fF_{zx} 
\left(
\frac{\partial \FF_{xu}}{\partial z} +
\frac{\partial \FF_{uz}}{\partial x} 
\right) +
\\
\fF_{xy} 
\left(
\frac{\partial \FF_{yu}}{\partial x} +
\frac{\partial \FF_{ux}}{\partial y} 
\right) +
\fF_{wx} 
\left(
\frac{\partial \FF_{uw}}{\partial x} +
\frac{\partial \FF_{xu}}{\partial w} 
\right) +
\\
\fF_{wy} 
\left(
\frac{\partial \FF_{uw}}{\partial y} +
\frac{\partial \FF_{yu}}{\partial w} 
\right) +
\fF_{wz} 
\left(
\frac{\partial \FF_{uw}}{\partial z} +
\frac{\partial \FF_{zu}}{\partial w} 
\right) +
\\
\FF_{ux} \frac{\partial \fF_{xu}}{\partial u} +
\FF_{uy} \frac{\partial \fF_{yu}}{\partial u} +
\FF_{uz} \frac{\partial \fF_{zu}}{\partial u} +
\FF_{uw} \frac{\partial \fF_{wu}}{\partial u} +
\\
\fF_{yz} \frac{\partial \FF_{yz}}{\partial u} +
\fF_{zx} \frac{\partial \FF_{zx}}{\partial u} +
\fF_{xy} \frac{\partial \FF_{xy}}{\partial u} +
\\
\fF_{wx} \frac{\partial \FF_{wx}}{\partial u} +
\fF_{wy} \frac{\partial \FF_{wy}}{\partial u} +
\fF_{wz} \frac{\partial \FF_{wz}}{\partial u} =
\\
\FF_{ux} \sF_x +
\FF_{uy} \sF_y +
\FF_{uz} \sF_z +
\FF_{uw} \sF_w.
\end{align*}
}

We must now consider the supplementary condition which relates \fF~ and \FF. If these two vectors are not equal, then the simplest form of a relation between them is 

\begin{equation}
\label{EQ:6}
\fF = \varepsilon \; \FF,
\end{equation}

where $\varepsilon$ is a five dimensional scalar which is some function of the physical state. Since the $w$-direction is a distinguished direction, we may have, instead of the five dimensional scalar $\varepsilon$, two quantities $\varepsilon_1$ and
$\varepsilon_2$, so that

\[
\fF_{ab} = \varepsilon_1 \; \FF_{ab} 
\]

where neither $a$ nor $b$ is equal to $w$, otherwise

\[
\fF_{wa} = \varepsilon_2 \; \FF_{wa}.
\]

In the world-surface $\varepsilon_1$ and $\varepsilon_2$ appear as four dimensional scalars. In the following we will maintain the simpler assumption (\ref{EQ:6}) and thus set $\varepsilon_1 = \varepsilon_2 = \varepsilon$.    
   
We have then

\begin{align*}
\fF_{yz} \frac{\partial \FF_{yz}}{\partial u} =
\frac{1}{2} \frac{\partial}{\partial u} \fF_{yz} \FF_{yz} - \frac{1}{2} \FF_{yz}^2 \frac{\partial \varepsilon}{\partial u},
\\
\FF_{ux} \frac{\partial \FF_{xu}}{\partial u} =
-\frac{1}{2} \frac{\partial}{\partial u}\fF_{ux} \FF_{ux} + \frac{1}{2} \fF_{ux}^2 \frac{\partial}{\partial u} \frac{1}{\varepsilon} =
\\
-\frac{1}{2} \frac{\partial}{\partial u}\fF_{ux} \FF_{ux} - \frac{1}{2} \FF_{ux}^2 \frac{\partial \varepsilon}{\partial u}
\\
\mbox{etc}.
\end{align*} 

After rearranging our long equation it becomes

\begin{equation}
\label{EQ:7}
\left\{
\begin{aligned}
-\left\{
\frac{\partial \GSF_{xu}}{\partial x} +
\frac{\partial \GSF_{yu}}{\partial y} +
\frac{\partial \GSF_{zu}}{\partial z} +
\frac{\partial \GSF_{wu}}{\partial w} +
\frac{\partial \GSF_{uu}}{\partial u}
\right\}
=
\\
\frac{1}{2} \sum \FF_{ab}^2 \frac{\partial \varepsilon}{\partial u} +
\sF_x \FF_{ux} +
\sF_y \FF_{uy} +
\sF_z \FF_{uz} +
\sF_w \FF_{uw}, 
\end{aligned}
\right.
\end{equation}

where the components of the five dimensional tensor \GSF~ are given by the expressions:

\begin{equation}
\label{EQ:8}
\left\{
\begin{aligned}
&\GSF_{xu} = 
\fF_{xy} \FF_{uy} +
\fF_{xz} \FF_{uz} +
\fF_{xw} \FF_{uw},
\\
&\GSF_{uu} = \frac{1}{2}
\{
\fF_{ux} \FF_{ux} +
\fF_{uy} \FF_{uy} +
\fF_{uz} \FF_{uz} +
\fF_{uw} \FF_{uw} -
\\
&\fF_{yz} \FF_{yz} -
\fF_{zx} \FF_{zx} -
\fF_{xy} \FF_{xy} -
\fF_{wx} \FF_{wx} -
\fF_{wy} \FF_{wy} -
\fF_{wz} \FF_{wz} 
\}.
\end{aligned}
\right.
\end{equation}

The equation (\ref{EQ:7}) expresses (when multiplied by $i c$) the energy theorem for the unified electromagnetic and gravitational field. We add to both sides of the equation the terms which refer to the matter stress-tensor, and obtain, when taking into consideration the expressions (\ref{EQ:5}) for the components of \TSF~ and the expressions (1) l.\ c.\ for the components of \FF,

\begin{equation}
\label{EQ:9}
\left\{
\begin{aligned}
&-\frac{\partial}{\partial x} \left\{ \GSF_{xu} + \TSF_{xu} \right\}
-\frac{\partial}{\partial y} \left\{ \GSF_{yu} + \TSF_{yu} \right\}
-\frac{\partial}{\partial z} \left\{ \GSF_{zu} + \TSF_{zu} \right\} -
\\
&-\frac{\partial}{\partial w} \left\{ \GSF_{wu} + \TSF_{wu} \right\}
-\frac{\partial}{\partial u} \left\{ \GSF_{uu} + \TSF_{uu} \right\} =
\\
&\frac{1}{2} \sum \FF_{ab}^2 \frac{\partial \varepsilon}{\partial u} +
\sF_x 
\left\{
\frac{\partial \varPhi_x}{\partial u} -\frac{\partial \varPhi_u}{\partial x} \right\} +
\sF_y 
\left\{
\frac{\partial \varPhi_y}{\partial u} -\frac{\partial \varPhi_u}{\partial y} \right\} +
\\
&\sF_z 
\left\{
\frac{\partial \varPhi_z}{\partial u} -\frac{\partial \varPhi_u}{\partial z} \right\} +
\sF_w 
\left\{
\frac{\partial \varPhi_w}{\partial u} -\frac{\partial \varPhi_u}{\partial w} \right\} +
\\
&\frac{\partial}{\partial x} \sF_x \varPhi_u +
\frac{\partial}{\partial y} \sF_y \varPhi_u +
\frac{\partial}{\partial z} \sF_z \varPhi_u +
\frac{\partial}{\partial w} \sF_w \varPhi_u +
\frac{\partial}{\partial u} \sF_u \varPhi_u -
\\
&\frac{1}{4}\frac{\partial}{\partial u}
\left\{
\sF_x \varPhi_x +
\sF_y \varPhi_y +
\sF_z \varPhi_z +
\sF_u \varPhi_u +
\sF_w \varPhi_w 
\right\}
\end{aligned}
\right.
\end{equation}

Both sides of this equation must be, thanks to the energy theorem (\ref{EQ:3}), equal to zero, and when we put the right hand equal to zero we obtain the condition imposed by the energy theorem on our theory. We note furhter that

\[
\varPhi_u
\left\{
\frac{\partial \sF_x}{\partial x} +
\frac{\partial \sF_y}{\partial y} +
\frac{\partial \sF_z}{\partial z} +
\frac{\partial \sF_u}{\partial u} +
\frac{\partial \sF_w}{\partial w} 
\right\} = 0,
\]

since the expression in the brackets is zero, which can be seen by differentiating equation (1) l.\ c.\ with respect to $x$, $y$ $z$, $u$, $w$, and adding. By setting the the right hand side of (\ref{EQ:9}) to zero yields thus

\begin{align*}
&\frac{1}{2} \sum \FF_{ab}^2 \frac{\partial \varepsilon}{\partial u} +
\sF_x \frac{\partial \varPhi_x}{\partial u} +
\sF_y \frac{\partial \varPhi_y}{\partial u} +
\sF_z \frac{\partial \varPhi_z}{\partial u} +
\sF_w \frac{\partial \varPhi_w}{\partial u} +
\sF_u \frac{\partial \varPhi_u}{\partial u} -
\\
&\frac{1}{4}\frac{\partial}{\partial u}
\left\{
\sF_x \varPhi_x +
\sF_y \varPhi_y +
\sF_z \varPhi_z +
\sF_u \varPhi_u +
\sF_w \varPhi_w 
\right\} = 0.
\end{align*}

This condition refers to the $u$-direction. The corresponding equations for the other directions are similar, and differ only in that the derivatives are with respect to some other coordinate. We may therefore summarize all five equations as

\begin{equation}
\label{EQ:10}
\left\{
\begin{aligned}
&\frac{1}{2} \sum \FF_{ab}^2 \cdot d\varepsilon +
\sF_x d\varPhi_x +
\sF_y d\varPhi_y +
\sF_z d\varPhi_z +
\sF_u d\varPhi_u +
\sF_w d\varPhi_w -
\\
&\frac{1}{4}
d(
\sF_x \varPhi_x +
\sF_y \varPhi_y +
\sF_z \varPhi_z +
\sF_u \varPhi_u +
\sF_w \varPhi_w
) = 0.
\end{aligned}
\right.
\end{equation}

When these conditions are satisfied, then the energy-momentum theorem is also valid in our theory\footnote{A comparison of the derivation of the equation (\ref{EQ:10}) with the corresponding considerations by \so{Mie} in Ann.\ d.\ Phys.\ \textbf{37}, p.\ 522 ff.\ and \textbf{40}, p.\ 29 ff.\ show in my opinion, that his conditions do not have the general validity for every theory as he claims in Phys.\ Zeischr. \textbf{10}, \ p. 175.}. If we assume that \sF~ and $\varPhi$ are related by equation (\ref{EQ:2a}), then we obtain from (\ref{EQ:10})

\[
\frac{1}{2} \sum \FF_{ab}^2 \cdot d\varepsilon +
\frac{1}{2} \beta d \sum \varPhi_a^2 - 
\frac{1}{4} d\left(\beta \sum \varPhi_a^2\right) = 0.
\]

We set

\begin{equation}
\label{EQ:11a}
\tag{11 a}
\left\{
\begin{aligned}
&\xi = \sum \varPhi_a^2,
\\
&\eta = 2 \sum \FF_{ab}^2,
\end{aligned}
\right.
\end{equation}

and obtain the conditions, after some simple rearranging and transformations, on the form

\begin{equation}
\label{EQ:12a}
\tag{12 a}
\xi d \beta - \beta d\xi = \eta d \varepsilon.
\end{equation}

These equations are valid only when we assume the relations (\ref{EQ:2a}). If instead we assume the more general relations (\ref{EQ:2}), we get in place of (\ref{EQ:12a})

\begin{equation}
\label{EQ:12}
\tag{12}
\xi_1 d \beta_1 + \xi_2 d \beta_2 - 
\beta_1 d\xi_1 - \beta_2 d\xi_2 = \eta d \varepsilon,
\end{equation}

where

\begin{equation}
\label{EQ:11}
\tag{11}
\left\{
\begin{aligned}
&\xi_1  = 
\varPhi_x^2 +
\varPhi_y^2 +
\varPhi_z^2 +
\varPhi_u^2,
\\
&\xi_2 = \varPhi_w^2,
\\
&\eta = 2 \sum \FF_{ab}^2.
\end{aligned}
\right.
\end{equation}

In the following we will investigate the equation (\ref{EQ:12a}) and assume that $\beta_1 = \beta_2$. Dividing by $\xi^2$ it becomes:

\[
d \, \frac{\beta}{\xi} = \frac{\eta}{\xi^2} \, d \varepsilon .
\]

In order to make the right hand side a complete differential, we will assume that $\varepsilon$ is only a function of $\frac{\eta}{\xi^2}$. Then $\frac{\beta}{\xi}$ becomes also a function of the same argument. Since $\varepsilon$ has the dimension of a pure number, we multiply the equation by a universal constant $\varkappa$ of such a dimension, that also

\[
\frac{\varkappa \eta}{\xi^2}
\]

becomes a pure number. Thus, $\varkappa$ has the dimension

\[
m \, l^3 \, l^{-2}
\]

that is, the square of an electric charge. The equation becomes now

\setcounter{equation}{12}

\begin{equation}
\label{EQ:13}
d\, \frac{\varkappa \beta}{\xi} = \frac{\varkappa \eta}{\xi^2} \, d\varepsilon.
\end{equation}

$\varepsilon$ and $\frac{\varkappa \beta}{\xi}$ are functions of $\frac{\varkappa \eta}{\xi^2}$. One may anticipate that, if our theory can  be developed in a satisfactory way, then the constant $\varkappa$ will assume an important role, determining the elementary charge $\pm e$, which is possible since $\varkappa$ has the same dimension as $e^2$.

\so{Planck}'s quantum of action $h$ is, as \so{Einstein} has remarked\footnote{\so{A. Einstein}, Phys.\ Zeitschr.\ \textbf{10}, p.\ 192, 1909.}, probably closely related to $e$ and has the same dimension as

\[
\frac{e^2}{c}
\]
 
(where $c$ is the velocity of light); that is, the same as that of $\varkappa/c$. Thus, through a satisfactory development of our theory one could hopefully determine $h$ in terms of $c$ and $\varkappa$. 

We have now to find the expressions which satisfy the conditions (\ref{EQ:13}). One possible and fairly plausible ansatz is the following one:

\begin{equation}
\label{EQ:14}
\left\{
\begin{aligned}
&\varepsilon - \varepsilon_0 = (n+1) \left( \frac{\varkappa \eta}{\xi^2} \right)^n,
\\
&\frac{\varkappa \beta}{\xi} = n \left( \frac{\varkappa \eta}{\xi^2} \right)^{n+1}. 
\end{aligned}
\right.
\end{equation}
 
$\varepsilon_0$ is a constant; however, to begin with we assume that $n$ is variable. When, for simplicity, we write

\begin{equation}
\label{EQ:15}
\frac{\varkappa \eta}{\xi^2} = \omega,   
\end{equation}

then we have

\begin{align*}
&\omega d\varepsilon = n(n+1) \omega^n d\omega + \omega^{n+1} dn + 
(n+1) \omega^{n+1} \ln \omega \, dn,
\\
&d\, \frac{\varkappa \beta}{\xi} = 
n(n+1) \omega^{n} d\omega + \omega^{n+1} dn + n \omega^{n+1} \ln \omega \, dn.
\end{align*}

We see that the condition (\ref{EQ:13}) is satisfied when, in every spacetime region where $\omega$ is neither equal to 0 nor 1, $n$ has a constant value. We have thus two possibilities: either is $n$ overall constant, or $n$ changes with discrete jumps on surfaces where $\omega = 0$ or $\omega = 1$. We will examine these two possibilities somewhat more closely.

When $n$ is overall constant, then one has not an absolutely empty space; where $\sum \FF_{ab}^2$ differs from zero there one has also electricity and matter, although their densities may be negligible small. The value of $n$ must be positive, furthermore we have to set $\varepsilon_0 = 1$, since for relatively weak fields, which we can study in the space which appears empty to us, $\varepsilon$ is practically constant and equal to one, while

\[
(n+1) \left( \frac{\varkappa \eta}{\xi^2} \right)^n 
\quad \mbox{small compared with} \quad \varepsilon_0 = 1.
\]

Only for the strong inter-atomic fields might $\varepsilon$ deviate noticeably from the value 1, and the densities of electricity and matter be significant. 

A seemingly more attractive assumption, than an overall constant $n$, is that $n$ changes with jumps when $\omega = 1$ and is equal to zero when $\omega < 1$. In this case one could make a principal distinction between aether (= empty space) and matter. One would have to set $\varepsilon_0 = 0$, and where $\omega < 1$ one would have $\varepsilon = \omega^0 = 1, \beta = 0$, so that in these regions there could be no electricity or matter (cmp. equation (\ref{EQ:2a})). When $\omega > 1$ one would have to assume a value different from zero for $n$, so that $\varepsilon$ differs from 1 and $\beta$ from 0. For $\omega > 1$ one could have, as an example, $n = 1$ or $n = \frac{1}{2}$.

In order to examine, to what extent the assumptions made above suffice for a theory of matter, we imagine an electrical particle (an electron), with radial symmetry, whose every part is at rest. Then 

\begin{equation}
\label{EQ:16}
\varPhi_x = \varPhi_y = \varPhi_z = 0. 
\end{equation}

$\varPhi_u$ and $\varPhi_w$, however, are functions of the distance $r$ from the center of the particle. We set

\begin{equation}
\label{EQ:17}
\varPhi_u = i \varPhi_e. 
\end{equation}

where $\varPhi_e$ is now the real, electrostatic potential. We obtain for the components \FF~ according to (1) l.\ c., since all the derivatives with respect to $u$ are zero,

\begin{equation}
\label{EQ:18}  
\left\{
\begin{aligned}
&\FF_{ux} = -i \frac{\partial \varPhi_e}{\partial x},\;  
\FF_{uy} = -i \frac{\partial \varPhi_e}{\partial y},\;
\FF_{uz} = -i \frac{\partial \varPhi_e}{\partial z},
\\
&\FF_{wx} = - \frac{\partial \varPhi_w}{\partial x},\;  
\FF_{wy} = - \frac{\partial \varPhi_w}{\partial y},\;
\FF_{wz} = - \frac{\partial \varPhi_w}{\partial z},
\\
&\FF_{xy} = \FF_{yz} = \FF_{zx} = 0.
\end{aligned}
\right.
\end{equation}

The components of \fF~ are obtained through multiplication by $\varepsilon$. From (\ref{EQ:16}) follows that

\begin{equation}
\label{EQ:19}
\sF_x = \sF_y = \sF_z = 0.
\end{equation}

Since all the field quantities depend only on the distance $r$, we have\footnote{Since the second term in the brackets dominate, $\eta$ is negative.}

\begin{equation}
\label{EQ:20}
\left\{
\begin{aligned}
&\eta = 2 \sum \FF_{ab}^2 = 
2 \left\{
\left( \frac{d \varPhi_w}{dr} \right)^2 - \left( \frac{d \varPhi_e}{dr} \right)^2
\right\},
\\
&\xi = \sum \varPhi_a^2 = \varPhi_w^2 - \varPhi_e^2.
\end{aligned}
\right.
\end{equation}  

If we denote the density of electricity by $\varrho$ and the density of the gravitating mass (denoted by $g \cdot \nu$ l.\ c.\ p.\ 4) by $\gamma$, then one has generally

\begin{equation}
\label{EQ:21}
\left\{
\begin{aligned}
&\sF_u = i \varrho,
\\
&\sF_w = -\gamma
\end{aligned}
\right.
\end{equation}  

and thus in our case

\begin{equation}
\label{EQ:22}
\left\{
\begin{aligned}
&\varrho = \beta \varPhi_e,
\\
&\gamma = - \beta \varPhi_w.
\end{aligned}
\right.
\end{equation}

The last two equations in (I) l.\ c.\ give when applying the theorem of \so{Gauss} to a thin spherical layer

\begin{equation}
\label{EQ:23}
\left\{
\begin{aligned}
&\frac{d}{dr} \left\{ 
r^2 \varepsilon \frac{d\varPhi_e}{dr}
\right\}
= -r^2 \beta \varPhi_e,
\\
&\frac{d}{dr} \left\{ 
r^2 \varepsilon \frac{d\varPhi_w}{dr}
\right\}
= -r^2 \beta \varPhi_w,
\end{aligned}
\right.
\end{equation}

The rest of the equations (I) and (II) l.\ c.\ are automatically satisfied.

When we insert the expressions for $\varepsilon$ and $\beta$ into (\ref{EQ:23}), we obtain a system of two second order differential equations, which determine $\varPhi_e$ and $\varPhi_w$ as functions of $r$. One easily finds the first integral, 

\[
r^2 \varepsilon \left\{
\varPhi_w \frac{d \varPhi_e}{dr} -
\varPhi_e \frac{d \varPhi_w}{dr}
\right\} = \mbox{const.},
\]

of the system. However, it is not possible to obtain a closed form of the determinate integral expressions for $\varPhi_e$ and $\varPhi_w$. To begin with there are moreover four undetermined constants of integration. Two of them can be determined by the fact that $\varPhi_e$ and $\varPhi_w$ are known for $r = \infty$. Since positive and negative electricity are present in equal amounts in the universe, we must have $\varPhi_e = 0$ for $r = \infty$. However, $\varPhi_w$ has for $r = \infty$ the big positive value

\[
\varPhi_{w \infty} = 0.984 \cdot 10^{24} \; \mbox{g}^{\frac{1}{2}} \; \mbox{cm}^{\frac{1}{2}} \; \mbox{sec}^{-1}.
\]

Furthermore, for large $r$ we should have the expansions  

\begin{equation}
\label{EQ:24}
\left\{
\begin{aligned}
&\varPhi_e = 0 + \frac{e}{4 \pi r} + \dots
\\
&\varPhi_w = \varPhi_{w \infty} - \frac{M}{4 \pi r} + \dots
\end{aligned}
\right.
\end{equation}

where $e$ is the elementary quantum of electricity (with positive or negative sign) and $M$ is the gravitating mass of the particle. $e$ and $M$ are the other two integration constants that cannot be determined from the theory developed. However, when these quantities are arbitrarily assigned, then it is impossible for our theory to lead to an atomistic structure for electricity and matter. We have thus to add new conditions to the foundations of the theory. It seems most expedient to assume two new algebraic conditions for the state variables $\xi$, $\eta$ at the center of the material particle (for example, it would always have to be $\xi = 0, \eta = 0$ at this point). We thus complete the theory by assuming that the center points of the electrons and atoms are singular points where the state variables are subject to conditions which do not apply to the rest of the field\footnote{The world-lines formed by these points are of course singular lines in the four dimensional world-surface.}. In the expansion (\ref{EQ:24}) for a particle at rest, the values $e$ and $M$ are no longer arbitrary, but are determined in an unique or multiple valued sense by some given system of two equations. The quantity $\varkappa$   will be included in the system of equations, and hence a close relation will be possible between the quantities $\varkappa$, $e^2$ and $M^2$, all which have the same dimension. 

The above considerations rest on the relations (\ref{EQ:12a}), which presume that $\beta_1 = \beta_2$. We can easily see that these assumptions lead to difficulties. Thus, we have for the particle in question

\begin{align*}
&e = \int \varrho \, dV = \int \beta \varPhi_e dV,
\\
&M = \int \gamma \, dV = - \int \beta \varPhi_w dV.
\end{align*}

$M$ is the gravitating mass of the particle, which is related through

\begin{equation}
\label{EQ:25}
M = \frac{c^2 m}{\varPhi_{w \infty}}
\end{equation}

to its inertial mass $m$\footnote{Cmp. \so{G. Nordstr{\"o}m}, Ann.\ d.\ Phys.\ p.\ 537, 1913.}. When $\varrho$ has the same sign in every part of the particle, then 

\[
\varPhi_e = \frac{1}{4 \pi} \int \frac{\varrho}{l} \, dV
\]

has the same sign as $\varrho$, and $\beta$ must be positive. Since $M$ is positive, $\varPhi_w$ must be predominantly positive. $\varPhi_w$ may well attain a negative value when large enough gravitating masses are involved and concentrated in small enough regions of space. This brings however with it other difficulties. When the gravitating mass of an electron, which is about

\[
M = 0.8 \cdot 10^{-30} \, \mbox{g}^{\frac{1}{2}} \, \mbox{cm}^{\frac{3}{2}} \, \mbox{sec}^{-1}, 
\]

is concentrated within a sphere of radius $a$, then we have at the surface of the sphere

\[
\varPhi_w = \varPhi_{w \infty} - \frac{M}{4 \pi a} \leq 0, 
\quad \mbox{when} \quad a \leq 6.4 \cdot 10^{-52} \; \mbox{cm}.
\]

If the main part of the electrical charge would be concentrated in such a small volume, then the electrostatic energy would be of the order

\[
\frac{e^2}{8 \pi a} = 1.7 \cdot 10^{32}\; \mbox{erg}
\]

corresponding to an inertial mass of the order

\[
\frac{1}{c^2} \cdot 1.7 \cdot 10^{32} = 1.9 \cdot 10^{11} \; \mbox{g}.
\] 

Since the correct value for $m$ is about $0.8 \cdot 10^{-27}$ g, one would have inside the electron a negative energy density of an enormous magnitude, which is quite unlikely.

The mentioned difficulties would hardly be less serious, if instead of the $\varrho$ assumed above we let its sign vary inside the particle.

On these grounds we can assert, without completing the integration of the differential equations (\ref{EQ:23}), that the theory based on the assumption $\beta_1 = \beta_2$ is unlikely to lead to the correct results.

When we drop the assumption $\beta_1 = \beta_2$, the situation improves, since $\beta_1$ can be positive while $\beta_2$ is negative, a desirable property as suggested by the equation

\begin{equation}
\label{EQ:26}
\left\{
\begin{aligned}
&\varrho = \beta_1 \varPhi_e,
\\
&\gamma = -\beta_2 \varPhi_w
\end{aligned}
\right.
\end{equation}

(which appears in place of (\ref{EQ:22})). When $\beta_1$ and $\beta_2$ are different quantities, then the number of independent quantities increases by one, and we may thus impose a new condition on them. The previous mathematical considerations are least affected if we set

\begin{equation}
\label{EQ:27}
\beta_1 \, \xi_1 = \beta_2 \, \xi_2.
\end{equation} 

Since $\xi_1$ is negative, while $\xi_2$ is positive, $\beta_1$ and $\beta_2$ will have opposite signs as desired. The equation (\ref{EQ:27}) implies that for a static field (cmp.\ (\ref{EQ:11}), (\ref{EQ:17}) and (\ref{EQ:26}))
 
\begin{equation}
\label{EQ:27a}
\tag{27 a}
\varrho \varPhi_e = \gamma \varPhi_w.
\end{equation}

That $\varrho \, \varPhi_e$ and $\gamma \, \varPhi_w$ are of the same order, independently of our assumptions, is shown by the following argument. We imagine that the electrical charge $e$ and the gravitating mass $M$ are uniformly distributed within a spherical volume of radius $a$. The densities of  electricity and gravitating mass are then of the orders

\[
(\varrho) = \frac{e}{a^3} \: \mbox{and} \: (\gamma) = \frac{M}{a^3}.   
\]

Inside the sphere $\varPhi_e$ is of the order 

\[
(\varPhi_e) = \frac{e}{a}.
\]

When $a$ is not too small, then $\varPhi_e$ is of the same order inside the sphere as at the infinity, and according to (\ref{EQ:25}) the gravitating mass $M$ is then of the order 

\[
(M) = \frac{c^2 m}{\varPhi_w}.
\]

The inertial mass $m$ is however, according to the common theory\footnote{Cmp. e.\ g.\ \so{M. Abraham}, Theorie der Elektrizit\"at II, p.\ 192 (Leipzig, 1905).} of the order

\[
(m) = \frac{e^2}{a c^2}.
\]

Since $\gamma$ is of the order 

\[
(\gamma) = \frac{e^2}{a^4 \varPhi_w},
\]

we see that according to the common theory $\varrho \, \varPhi_e$ and $\gamma \, \varPhi_w$ are of the same order, which lends some support to the assumption (\ref{EQ:27}).

We turn next to the theory based on the ansatz (\ref{EQ:27}). The equation yields by differentiation

\[
\xi_1 d\beta_1 - \beta_2 d\xi_2 =
\xi_2 d\beta_2 - \beta_1 d\xi_1,
\]

and from the equation (\ref{EQ:12}) p.~\pageref{EQ:12}, on which we will now base our considerations, we obtain

\[
2\xi_1 d\beta_1 - 2\beta_2 d\xi_2 =
2\xi_2 d\beta_2 - 2\beta_1 d\xi_1 = \eta d\varepsilon.
\]   

Through dividing by $2 \xi_1 \xi_2$ we obtain, since

\begin{gather*}
\frac{\beta_1}{\xi_2} = \frac{\beta_2}{\xi_1},
\\
\frac{d\beta_1}{\xi_2} - \frac{\beta_1 d\xi_2}{\xi_2^2} =
\frac{d\beta_2}{\xi_1} - \frac{\beta_2 d\xi_1}{\xi_1^2} =
\frac{\eta}{2 \xi_1 \xi_2} d\varepsilon,
\end{gather*}

and in place of (\ref{EQ:13}) we obtain the two equations

\begin{equation}
\label{EQ:28}
\left\{
\begin{aligned}
d \, \frac{\varkappa \beta_1}{\xi_2} = \acute{\omega} d\varepsilon,
\\
d\, \frac{\varkappa \beta_2}{\xi_1} = \acute{\omega} d\varepsilon.
\end{aligned}
\right.
\end{equation}

Here

\begin{equation}
\label{EQ:29}
\frac{\varkappa \eta}{2 \xi_1 \beta_2} = \acute{\omega},
\end{equation}

and $\varkappa$ is as before a universal constant of the dimension of the square of electric charge.

In analogy with the equation (\ref{EQ:13}) the equations (\ref{EQ:28}) have a solution of the form

\begin{equation}
\label{EQ:30}
\left\{
\begin{aligned}
&\varepsilon - \varepsilon_0 = (n+1) \, \acute{\omega}^n,
\\
&\frac{\varkappa \beta_1}{\xi_2} = \frac{\varkappa \beta_2}{\xi_1} =
n \, \acute{\omega}^{n+1}. 
\end{aligned}
\right.
\end{equation}

$n$ is again overall constant, or changes by jumps at surfaces where $\acute{\omega} = 1$.

When we consider as on p.~\pageref{EQ:19} a particle at rest with spherical symmetry, we have 

\begin{equation}
\label{EQ:31}
\left\{
\begin{aligned}
&\xi_1 = -\varPhi_e^2 , \: \xi_2 = \varPhi_w^2,
\\
&\eta = 
2 \left\{
\left( \frac{d \varPhi_w}{dr} \right)^2 - \left( \frac{d \varPhi_e}{dr} \right)^2
\right\},
\end{aligned}
\right.
\end{equation}  

and in place of equation (\ref{EQ:23}) we obtain the following:

\begin{equation}
\label{EQ:32}
\left\{
\begin{aligned}
&\frac{d}{dr} \left( 
r^2 \varepsilon \frac{d\varPhi_e}{dr}
\right)
= -r^2 \beta_1 \varPhi_e,
\\
&\frac{d}{dr} \left( 
r^2 \varepsilon \frac{d\varPhi_w}{dr}
\right)
= -r^2 \beta_2 \varPhi_w.
\end{aligned}
\right.
\end{equation}  

By inserting the expressions for $\beta_1$, $\beta_2$
and $\varepsilon$ one obtains, as  before, two differential equations for
$\varPhi_e$ and $\varPhi_w$, which determine them as functions of $r$. For large $r$ we have as on p.~\pageref{EQ:24} the expansions

\begin{equation*}
\left\{
\begin{aligned}
&\varPhi_e = 0 + \frac{e}{4 \pi r} + \dots
\\
&\varPhi_w = \varPhi_{w \infty} - \frac{M}{4 \pi r} + \dots
\end{aligned}
\right.
\end{equation*}

In order that the theory indeed should determine the structure of electricity and matter, $e$ and $M$ cannot be arbitrary constants of integration, and this is achieved when we consider the center points of electrons and atoms to be  singular points of the field, where the values of the state variables $\xi_1$, $\xi_2$, $\eta$ are subject to two algebraic conditions. Thus would $e$ and $M$ be determined in an unique or multiple valued sense by two equations containing $\varkappa$.

How to formulate these special conditions at the singular points, is a question whose answer requires the integration of the two differential equations (\ref{EQ:32}). Since this is beyond our means, there remains a large gap in our theory.

We will consider some details of the theory. With regard to $n$ the most plausible values\footnote{Also the quite special case $n = -1$ is to be considered. This would give for $\varepsilon$ the constant value $\epsilon_0$ (= 1).} in case of an overall constant value are $n$ = 1 and $n$ = $\frac{1}{2}$. 

However, as has been pointed out before, $n$ may change by jumps at surfaces where $\acute{\omega}$ = 1. We will examine whether this case can be attained in a plausible way. According to the common theory\footnote{Cmp. e.\ g.\ \so{M. Abraham}, Theorie der Elektrizit\"at II, p.\ 193 (Leipzig, 1905).} the electron radius is of the order of

\[
a = 10^{-13} \; \mbox{cm}.
\]

For this value for $a$ the field attains at the electron surface:

\begin{align*}
\varPhi_e &= \frac{e}{4 \pi \cdot 10^{-13}} = 1.3 \cdot 10^3,
\\
\frac{d \varPhi_e}{dr} &= -\frac{e}{4 \pi \cdot 10^{-26}} = - 1.3 \cdot 10^{16}.
\end{align*}
   
(The units are g, cm, sec.) We thus find that $\eta$ is of the order of $-10^{32}$ at the electron surface, $\xi_1$ is of the order of $-10^{6}$, and $\xi_2$ (everywhere) of the order of $10^{48}$. In order that $\acute{\omega}$ = 1, $\varkappa = \frac{2 \xi_1 \xi_2}{\eta}$ must be of the order of $10^{22}$. Then, however, a close connection between $\varkappa$ and

\[
e^2 = 4 \pi \cdot (4.65)^2 \cdot 10^{-20} 
\]

becomes impossible. We won't easily give up the requirement that $\varkappa$ should determine the elementary quantum of charge, but we have to state that the assumption of a discontinuous change in $n$ (where $n$ = 0 for $\acute{\omega} < 1$), though having desirable consequences, also brings with it severe difficulties. One could hope that the assumption of an overall constant $n$ would show greater promise for the theory. Unfortunately, the difficulties associated with the integration of the differential equations (\ref{EQ:32}) pose great obstacles for a further development of the theory. 

As a brief summary we can say, that we have succeeded in setting up general field equations for the total field, in which there appears a universal constant of the dimension of the square of the electrical charge (or, whose square divided by $c$ equals \so{Planck}'s quantum of action). It seems though that the atomistic structure of electricity and matter does not emerge from the theory, whence it is assumed that the center points of atoms and electrons are singular points, subject to special conditions. Then the theory can explain the existence of discrete elementary particles; however, the important question, whether an appropriate selection of the undetermined constants and conditions also may lead to quantitatively correct results, remains unresolved due to mathematical difficulties. 
\\[10pt] {\hspace*{20pt}\so{Helsingfors}, May 1915.}

\end{document}